\documentclass[prl,aps,twocolumn,preprintnumbers,nofootinbib,notitlepage,showkeys]{revtex4}
\usepackage{amssymb, amsthm}
\usepackage{amsmath}
\usepackage{graphicx}
\usepackage{hyperref}
\usepackage{dcolumn}
\usepackage{bm}%
\usepackage{float} 
\begin{document}

\title{Theory Driven Evolution of the Weak Mixing Angle}%

\author{Jens Erler$^1$, Rodolfo Ferro-Hernández$^1$ and Simon Kuberski$^2$}\affiliation{$^1${\rm PRISMA}$^+$ Cluster of Excellence and 
Institute for Nuclear Physics, Johannes Gutenberg-University, 55099 Mainz, Germany\\
$^2$Department of Theoretical Physics, CERN, 1211 Geneva 23, Switzerland}

\date{\today}

\begin{abstract}
We present the first purely theoretical calculation of the weak mixing angle in the $\overline{\mathrm{MS}}$ scheme at low energies 
by combining results from lattice QCD with perturbation theory. 
We discuss its correlation with the hadronic contribution to the anomalous magnetic moment of the muon and 
to the energy dependence of the electromagnetic coupling. 
We also compare the results with calculations using cross-section data as input.
Implications for the Standard Model prediction of the mass of the $W$ boson are also discussed.
\end{abstract}

\keywords{Electroweak Precision Tests, Parity Violation, Lattice QCD, Renormalization.}
\preprint{MITP-24-051, CERN-TH-2024-071}
\maketitle

The weak mixing angle, $\theta_W$, is a central parameter in the Standard Model (SM) of particle physics, featuring prominently in many precision observables,
such as parity violation, neutrino physics, or $Z$-pole measurements. 
Thus, it serves as a useful tool for studying the consistency of the SM across different energy scales.
In particular, the upcoming low-energy parity violating electron scattering experiments P2 at Mainz~\cite{Becker:2018ggl} and MOLLER~\cite{MOLLER:2014iki} 
at JLab profit from an enhanced sensitivity due to an accidental suppression of the left-right polarization cross section asymmetries  
which are proportional to $1 - 4 \sin^2\theta_W \ll 1$.  
Just as the predecessor experiments SLAC--E158~\cite{SLACE158:2005uay} and JLab--Qweak~\cite{Qweak:2018tjf} they 
are sensitive to higher-order SM corrections~\cite{Czarnecki:1995fw,Czarnecki:2000ic}, especially from the $\gamma Z$ vacuum polarization function.
The level of precision at P2 and MOLLER requires even the inclusion of two-loop electroweak effects~\cite{Du:2019evk,Erler:2022ckm}. 

Analogous to the dependence on the energy scale $\mu$ of the electromagnetic coupling, $\hat{\alpha}(\mu)$, {\em i.e.\/}, its {\em running}, higher order terms, 
in particular $\gamma Z$ vacuum polarization effects, can be incorporated into the running of the weak mixing angle. 
The large logarithms that emerge when using the weak mixing angle measured at high energy colliders as input in low-energy processes, 
call for a systematic inclusion and re-summation of higher order corrections. 
This procedure is renormalization scheme dependent. 
For computational simplicity we choose the $\overline{\mathrm{MS}}$ scheme (denoted by a caret), 
where $\hat\theta_W$ is defined in terms of the SM gauge couplings $\hat g$ and $\hat g'$,
\begin{equation}
\hat{s}^2\equiv\sin^2\hat{\theta}_W=\frac{\hat g^{\prime2}}{\hat g^2 + \hat g^{\prime 2}}\ .
\end{equation}
Ref.~\cite{Erler:2004in} derived a relation between $\hat\theta_W(\mu)$ and $\hat{\alpha}(\mu)$ in the $\overline{\mathrm{MS}}$ scheme 
(see Ref.~\cite{Jegerlehner:1985gq} for an approach using a different scheme). 
This allowed to straightforwardly include non-perturbative hadronic contributions to the running of $\hat\theta_W$ 
by using $e^+e^-$ annihilation data in a dispersion relation. 
The main source of uncertainty was due to the necessary flavor separation, defined as the contribution of the strange quark current relative to the up and down 
quark currents. 
Significant progress was reported in Ref.~\cite{Erler:2017knj} where improved data, a more precise flavor separation estimate, 
and the inclusion of higher-order perturbative QCD (pQCD) corrections, led to a noticeable reduction in the total uncertainty in $\hat{s}^2(0)$. 

In a different development, there is a discrepancy between the measured~\cite{Muong-2:2006rrc,Muong-2:2021ojo,Muong-2:2023cdq} and 
predicted~\cite{Aoyama:2020ynm,Keshavarzi:2019abf,Davier:2019can,Jegerlehner:2019lxt} values of the anomalous magnetic moment 
of the muon\footnote{To find details about all the contributions to $a_{\mu}$, refer to Ref.~\cite{Aoyama:2020ynm} and the cited references therein.}, $a_\mu$ (employing $e^+e^-$ data for the hadronic vacuum polarization contribution $a^{\mathrm{hvp}}_\mu$).
Recently, the results of the CMD-3 experiment~\cite{CMD-3:2023alj} revealed further tension when compared with $e^+e^-$ data sets from previous experiments. 
Moreover, ab-initio lattice QCD (LQCD) calculations~\cite{Giusti:2019xct, Shintani:2019wai, FermilabLattice:2019ugu, Gerardin:2019rua, Borsanyi:2020mff, Aubin:2022hgm}  are in reasonable agreement with the measured $a_\mu$ and CMD-3. 
Since $e^{+}e^{-}$ data also enter into calculations of $\hat{\alpha}$ and $\hat\theta_W$, these tensions should affect these quantities as well, 
and an effort is required to incorporate LQCD results in the respective SM predictions.  
A first step in this direction~\cite{Erler:2023hyi} showed how LQCD can be used in an optimal way to include hadronic effects into the running of $\hat{\alpha}$.
A parametric expression in terms of input parameters was given, simplifying the implementation into global electroweak (EW) fits. 
 
Applying the framework of Refs.~\cite{Erler:2004in,Erler:2017knj,Erler:2023hyi}, and obtaining the flavor separation entirely from LQCD, 
we derive a purely theoretical SM prediction of $\hat{s}^2(0)$ in terms of $\hat{s}^2(M_Z)$, where $M_Z$ is the mass of the $Z$ boson.
We also give a simplified formula which can be included into EW fitting libraries.
Finally, we quantify the correlations between $\hat{s}^2(0), \hat{\alpha}(M_Z)$, and $a^{\mathrm{hvp}}_\mu$.

We find a discrepancy between the lattice and the data-driven predictions for the running. 
This tension (which induces a positive shift of $\sim 8 \times 10^{-5}$ in $\hat{s}^2(0)$ when experimental data are replaced by LQCD), 
is about 30\% of the uncertainty anticipated for future low-energy parity-violating experiments. 
On the other hand, if this issue can be resolved, we would be left with a residual uncertainty of $\delta\hat{s}^2(0)=2\times 10^{-5}$, 
negligible for the low-energy parity-violating experiments in the foreseeable future.

Our starting point is the vacuum polarization function, 
\begin{equation}
\hat{\Pi}(q^2,\mu^2)=-\frac{i}{3q^2}\int d^4x e^{iqx}\langle 0|TJ^\mu(x) J_{\, \mu}(0)| 0\rangle
\end{equation}
where $J^\mu$ is the electromagnetic current. 
LQCD computes the subtracted vacuum polarization function, $\Pi(-Q^2) = \hat{\Pi}(0,\mu^2)-\hat{\Pi}(-Q^2,\mu^2)$. 
For large enough~$Q^2$, pQCD can be used to obtain the subtraction constant $\hat{\Pi}(0,\mu^2)$ 
which encodes the running of the $\overline{\mathrm{MS}}$ couplings. 
Indeed, setting $Q^2 = \mu^2$, we arrive at the result in the $\overline{\mathrm{MS}}$ scheme by adding 
\begin{equation}
\hat{\Pi}(-Q^2,Q^2)=\sum_{f}\frac{Q^2_f}{4\pi^2}\sum^{3}_{n=0}c_n \frac{\hat{\alpha}_s^n (Q^2)}{\pi^n} ,\label{eq:Changescheme}
\end{equation}
to the lattice results.
Here, $\hat\alpha_s$ is the strong coupling, and the constants,
\begin{align*}
c_0 &= \frac{5}{3} - 6\frac{\hat{m}^2_f}{Q^2}\ , &
c_1 &= -0.22489 - 16\ \frac{\hat{m}^2_f}{Q^2}\ , \\
c_2 &= 0.8522 - 144.85\ \frac{\hat{m}^2_f}{Q^2}\ , &
c_3 &= 5.588\ ,
\end{align*}
where $\hat{m}_f$ is the $\overline{\mathrm{MS}}$ mass of fermion $f$ at the scale $Q$, were obtained with the help of 
Refs.~\cite{Nesterenko:2016pmx, Maier:2011jd, Chetyrkin:1996cf, Chetyrkin:2000zk, Shifman:1978bx, Eidelman:1998vc, Surguladze:1990sp, Braaten:1991qm}.
We use this conversion formula only at energy scales where the three light quarks can be treated as approximately degenerate, 
so that the disconnected piece\footnote{Contribution from two closed fermion loops coupled to the EW currents
which is proportional to $(\sum_f Q_f)^2$ rather than $\sum_f Q_f^2$.} vanishes.  

The running of $\hat{\alpha}$ is given by
\begin{equation}
\hat{\alpha}(\mu)=\frac{\alpha}{1-\Delta\hat{\alpha}(\mu)}\ ,
\label{eq:alpharun}
\end{equation}
where 
$\Delta\hat{\alpha}(\mu) = 4 \pi\alpha\hat{\Pi}(0,\mu^2)$ and $\alpha \approx 1/137.036$. 
In a first step and with the help of Eqs.~(\ref{eq:Changescheme}) and (\ref{eq:alpharun}), 
we compute $\hat{\alpha}(\mu)$ at some reference scale $\mu$ somewhat above the hadronic region employing LQCD from the Mainz collaboration \cite{Ce:2022eix}.
Then, we use the renormalization group equation (RGE) which is known up to order $\hat{\alpha}^4_s$~\cite{Baikov:2012zm} to compute\footnote{The 
differential equation can be solved iteratively in analytical form~\cite{Erler:1998sy} or numerically~\cite{Erler:2023hyi}.
The numerical difference which is related to the unknown perturbative orders is negligible.} $\hat\alpha(M_Z)$. 
At the threshold of each particle, the matching conditions given in Refs.~\cite{Chetyrkin:1997un, Sturm:2014nva} are applied.  
Thus, dependence on $\hat{\alpha}_s$ and the heavy quark masses, $\hat{m}_c$, and $\hat{m}_b$, is induced.

\begin{table}[t]
\centering
\begin{tabular}{|l|c|lll|}
\hline
\rule{0pt}{2.5ex} parameter~ & ~result $\times 10^4$~ & \multicolumn{3}{c|} {correlations} \\
\hline
\rule{0pt}{2.5ex}  ~~~~$\Pi_{\mathrm{disc}}$ & \phantom{.}$ -3.8 \pm 0.2$ & ~1.0 & ~~0.8 & ~0.8~ \\
\rule{0pt}{0ex} ~~~~$\Pi_s$ & $\phantom0 83.0 \pm 1.4$ & ~0.8 & ~~1.0 & ~0.96~ \\
\rule{0pt}{0ex} ~~~~$\Pi_{ud}$ & $587.8 \pm 8.3$            & ~0.8 & ~~0.96 & ~1.0 \\
\hline
\end{tabular}
\caption{Values, errors, and correlations for the vacuum polarization function at $Q^2 = 4 \mbox{ GeV}^2$. 
They were obtained from Appendix F of Ref.~\cite{Ce:2022eix} assuming that the disconnected contribution is associated with the $u$ and $d$ quarks. The isospin error was assigned entirely to the up and down contribution.}
\label{tab:errorslattice}
\end{table}

Since the running of $\hat\theta_W$ is related to that of $\hat{\alpha}$ through the photon vector polarization function, 
we can relate the solutions to their RGEs~\cite{Erler:2004in,Erler:2017knj}, and in the process re-sum the logarithms in $\hat s^2(\mu)$,
\begin{align} 
&\hat s^2(\mu) = \hat s^2(\mu_0) \frac{\hat\alpha(\mu)}{\hat\alpha(\mu_0)} + \lambda_1 \left[ 1 - \frac{\hat\alpha(\mu)}{\hat\alpha(\mu_0)} \right] + \nonumber \\[6pt]
&\frac{\hat\alpha(\mu)}{\pi} \left[ \frac{\lambda_2}{3} \ln\frac{\mu^2}{\mu_0^2} + \frac{3\lambda_3}{4} \ln \frac{\hat\alpha(\mu)}{\hat\alpha(\mu_0)} + 
\tilde\sigma(\mu_{0}) - \tilde\sigma(\mu) \right]\ .
\label{eq:MASTEREQUATION}
\end{align}
The $\lambda_i$ are constants, and the quantities $\tilde{\sigma}$ contain the contributions from disconnected diagrams.
Both depend on the number of active particles $n_f$, so that $\hat{s}^2(\mu)$ is a piecewise function in which the number of particle types change when a threshold is 
crossed and the matching conditions~\cite{Erler:2004in,Erler:2017knj} are used. 
With $\hat{\alpha}(\mu)$ known at the reference scale $\mu$, we use Eq.~(\ref{eq:MASTEREQUATION}) and the matching conditions 
to compute $\hat{s}^2(\mu)$ in terms of $\hat{s}^2(M_Z)$.

However, the aforementioned dependence of Eq.~(\ref{eq:MASTEREQUATION}) on $n_f$ requires separate information regarding the relative 
contributions of strange {\em versus\/} up and down quarks (flavor separation) in the hadronic (non-perturbative) region, as well as from disconnected diagrams. 
To address this, we translate the results by the Mainz lattice collaboration~\cite{Ce:2022eix}, which are given in terms of the $SU(3)$ labeled vacuum polarization 
functions\footnote{In principle, different disconnected contributions enter into $\Pi_{08}$ and $\Pi_{88}$ which would present us with four unknowns for three equations. 
But by noticing that the disconnected part is mainly due to the up and down quarks (as can be verified by lattice data at physical quark mass, see also the results compiled in Table~4 of Ref.~\cite{Ce:2022eix}),
we can solve the system.}, 
$\Pi_{33}$, $\Pi_{88}$, and $\Pi_{08}$, into the connected pieces $\Pi_{\mathrm{ud}}$ and $\Pi_{s}$, as well as the disconnected piece  $\Pi_{\mathrm{disc}}$.
This is shown in Tab.~\ref{tab:errorslattice} together with the associated correlation matrix which we computed by assuming that each lattice error induced by a given source 
(like scale setting, model error or statistical) enters fully correlated\footnote{We based our assumptions on the errors reported in Ref.~\cite{Ce:2022eix} for the linear combinations $\Pi_{\gamma\gamma}$ and $\Pi_{\gamma Z}$. Ultimately, the running of $\hat{s}^2$ is controlled by $\Pi_{\gamma Z}$ (up to re-summations computed in this study). Ignoring this effect, the final error on the running from LQCD corresponds to that of $4\pi\alpha\Pi_{\gamma Z}$. Therefore a change in our assumptions changes marginally the results of this paper, since our assumptions reproduce the error on $\Pi_{\gamma Z}$.}. 
Finally, we use Eq.~(\ref{eq:Changescheme}) to convert these results to the $\overline{\mathrm{MS}}$ scheme\footnote{The use of pQCD down to scales of 
$\mu \sim 2$~GeV can be justified by the recent analyses in Refs.~\cite{Jegerlehner:2019lxt,Davier:2023hhn,Hernandez:2023ipz}.}, and with the flavor separation 
at hand\footnote{For more technical steps on how Eq.~(\ref{eq:MASTEREQUATION}) is used given a known flavor separation, see \cite{Erler:2017knj,Erler:2004in}.}
we can apply Eq.~(\ref{eq:MASTEREQUATION}). 

\begin{table}[t]
\centering
\begin{tabular}{|l|c|}
\hline 
\rule{0pt}{2.5ex} source & ~$\delta\sin^2\hat{\theta}_{W}(0)\times10^{5}$~ \\
\hline 
\rule{0pt}{2.5ex} LQCD & 2.3~ \\
\rule{0pt}{0ex} pQCD & 0.1~ \\
\rule{0pt}{0ex} condensates~ & 0.2~ \\
\hline 
\rule{0pt}{2.5ex} total & 2.3~ \\
\hline 
\end{tabular}
\caption{Uncertainties in the calculation of the low energy weak mixing angle.}
\label{tab:errors}
\end{table}

We computed $\hat{s}^2(0)$ numerically, displaying explicitly the dependence on $\hat{s}^2(M_Z)$, $\hat\alpha_s$, $\hat m_c$, $\hat m_b$, the LQCD input, 
as well as the strange quark and gluon condensates\footnote{The condensates in the last two terms arise~\cite{Erler:2023hyi}  from 
the operator product expansion of the scheme conversion formula (\ref{eq:Changescheme}).}.
We write our main result as $\hat{s}^2(0) = \hat{\kappa}(0)\hat s^2(M_Z)$, with
\begin{align}
    \hat{\kappa}(0)&= 1.03234 - 0.43\, \Delta\hat s^2_Z + 0.030\, \Delta\hat\alpha_s \nonumber \\[3pt] 
    &- 0.0012\, \Delta\hat m_c - 0.0003\, \Delta\hat m_b \nonumber \\[3pt]
    &- 0.111\, \Delta\Pi_{\mathrm{disc}} + 0.206\, \Delta\Pi_s + 0.087\, \Delta\Pi_{ud} \nonumber \\[3pt]
    &+ \frac{0.003}{\mathrm{GeV}^4}\, \langle m_s \bar{s}s \rangle + \frac{0.0004}{\mathrm{GeV}^4}\, \left\langle \frac{\hat\alpha_s}{\pi}G^2 \right\rangle\ ,
\label{eq:kappa}
\end{align} 
where we defined,
\begin{align}
\Delta\hat{s}^2_Z &\equiv \hat{s}^2(M_Z) - 0.23122\ , \nonumber \\
\Delta\hat{\alpha}_s &\equiv \hat{\alpha}_s(M_Z) - 0.1185\ , \nonumber \\
\Delta\hat{m}_c &\equiv \hat{m}_c(\hat{m}_c) - 1.274 \mbox{ GeV}\ , \nonumber \\
\Delta\hat{m}_b &\equiv \hat{m}_b(\hat{m}_b) - 4.180 \mbox{ GeV}\ .
\label{eq:parameters}
\end{align} 
$\Delta\Pi_X$ is the difference between $\Pi_X$ ($X = \mathrm{disc}, s$, or $ud$) at $Q^2 = 4 \mbox{ GeV}^2$ and the central value shown in Tab.~\ref{tab:errorslattice}.

\begin{figure}[t]
    \centering
    \includegraphics[width=0.48\textwidth]{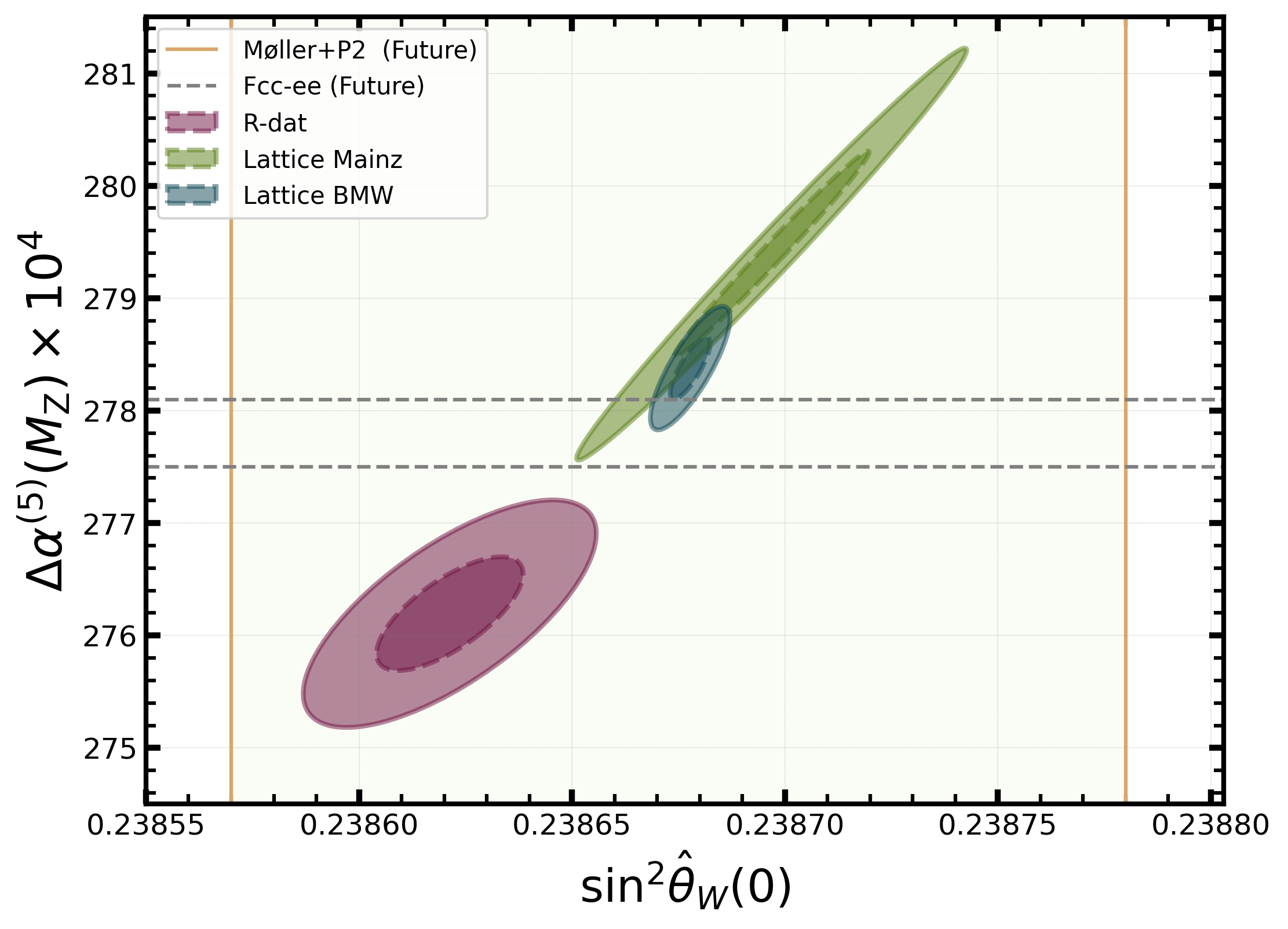}
    \caption{$\Delta\chi^2 = 1$ and $\Delta\chi^2 = 4$ contours for $\sin^2\hat{\theta}_{W}(0)$ and $\Delta\alpha^{(5)}(M_Z)$. 
    The yellow region represents the expected combined 1~$\sigma$ band for P2 and MOLLER. 
    The horizontal grey dashed band shows a projection~\cite{Blondel:2021ema} for the FCC-ee. 
    Only low energy (data and lattice) and pQCD errors are included.}
    \label{fig:confidence}
\end{figure}

Eq.~(\ref{eq:kappa}) shows that the LQCD uncertainty amounts to $1.0\times10^{-4}$ in $\hat{\kappa}(0)$ and the perturbative uncertainty, conservatively taken to correspond to the last known terms in the RGE (of order $\hat{\alpha}^4_s$), the decoupling relations, 
and the scheme conversion in Eq.~(\ref{eq:Changescheme}), is $4\times10^{-6}$.
The uncertainty from the condensates amount to $1\times10^{-5}$ (taking a conservative 100\% error of $0.003 \mbox{ GeV}^4$ 
in the strange quark condensate~\cite{McNeile:2012xh} and of $0.01\,\mathrm{GeV}^4$ in the gluon condensate~\cite{Narison:2011xe,Dominguez:2014fua}).  
The corresponding error budget for $\hat{s}^2(0)$ is shown in Tab.~\ref{tab:errors}. 
The linearized result~(\ref{eq:kappa}) approximates the exact numerical solution to better than 1~ppm even for values $3~\sigma$ away from the reference values.  

One can compare these results to those that use $e^{+}e^{-}\rightarrow\mathrm{hadrons}$ cross section data as input. 
There, large terms proportional to powers of $\pi^2$ are introduced when passing from timelike to spacelike momenta, enhancing the pQCD contribution and uncertainty. 
Furthermore, the result from the data driven approach\footnote{Due to different reference values,
the central value in~Ref.\cite{Erler:2017knj} is slightly different.}, $\hat{\kappa}(0)_{e^{+}e^{-}} = 1.03201\pm0.00008$~\cite{Erler:2017knj}, 
differs from Eq.~\eqref{eq:kappa} by $0.00033$ or about $3~\sigma$.
This is another reflection of the tension between LQCD and cross section data observed in the context of $\hat{\alpha}$ and $a_\mu$.

\begin{figure}[t] 
    \centering
    \includegraphics[width=0.47\textwidth]{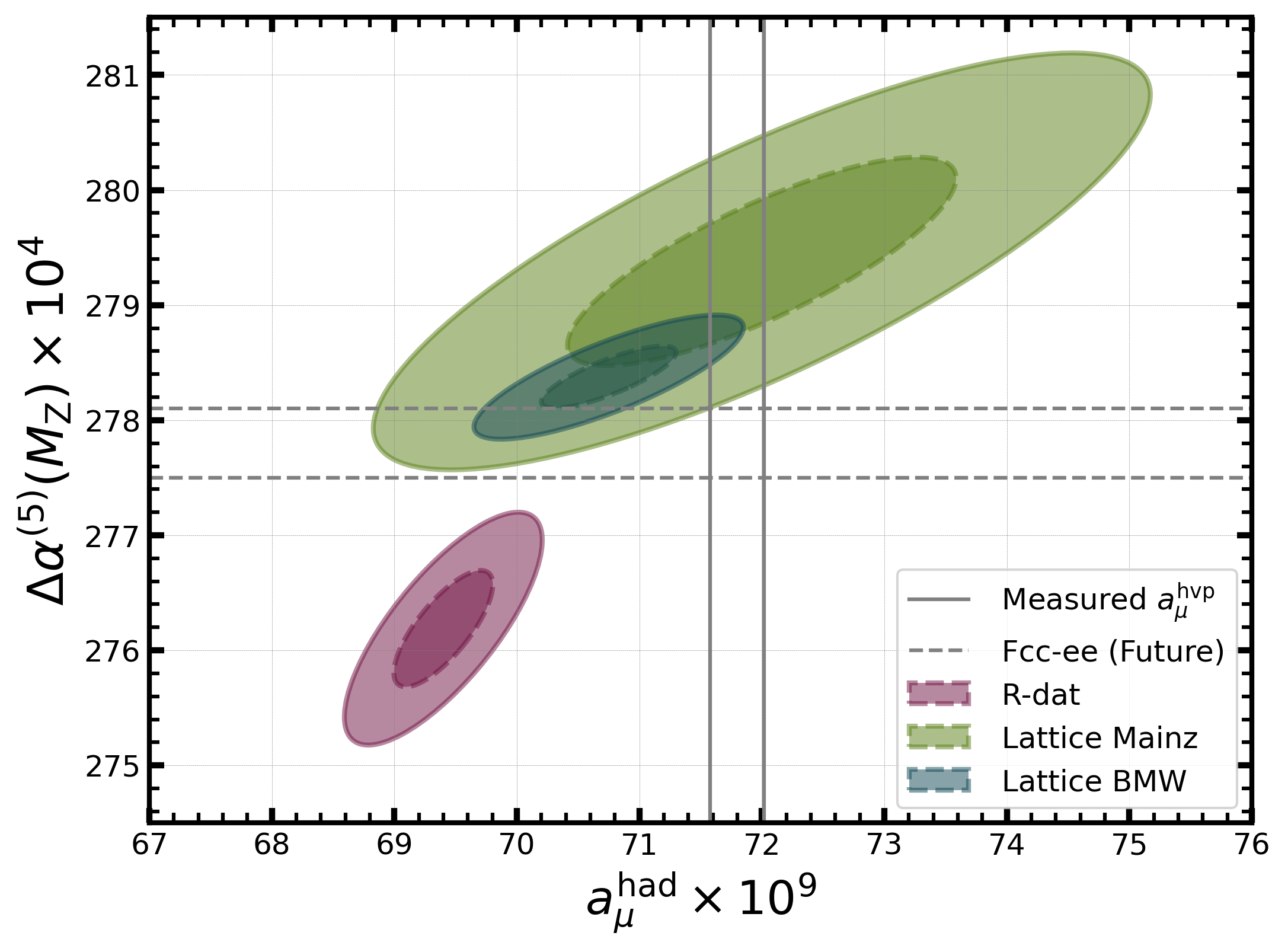}
    \caption{$\Delta\chi^2 = 1$ and $\Delta\chi^2 = 4$ contours for $\Delta\alpha^{(5)}(M_Z)$ and $a^{\mathrm{hvp}}_\mu$ ({\em cf.\/} Fig.~\ref{fig:confidence}).
    The width of the vertical band corresponds to the current experimental uncertainty in $a_\mu$.}
    \label{fig:confidenceamuda}
\end{figure}

As shown in Fig.~\ref{fig:confidence}, $\hat{s}^2(0)$ and $\hat{\alpha}$ are correlated.
Three input cases are considered for illustration, namely from LQCD~\cite{Ce:2022eix}, from $e^{+}e^{-}$ data~\cite{Davier:2019can} and from 
the lattice BMW collaboration~\cite{Budapest-Marseille-Wuppertal:2017okr}, where we assumed the same correlation between flavors as at Mainz. 
 
We also show the $1~\sigma$ bands projected for the future parity violation measurements~\cite{Becker:2018ggl,MOLLER:2014iki} (vertical yellow band) 
and the FCC-ee~\cite{Blondel:2021ema} (horizontal dashed grey line). 
In order not to dilute the tension between LQCD and data, the dependence on $\hat{\alpha}_s$ and the $\hat m_q$ is ignored in all figures.
A larger correlation is seen when LQCD results are used, as unlike in the data driven approach the flavor separation uncertainty entering in $\hat{s}^2(0)$ is correlated 
with the uncertainty in the sum of all flavors.

Investigating this kind of theoretical correlation is particularly important for EW global fits in the ultra-precision era.
Thus, the correlations of $\Delta\alpha^{(5)}(M_Z)$ and $\hat{s}^2(0)$ with $a^{\mathrm{hvp}}_{\mu}$ need to be evaluated and implemented, as well.
Since the calculation of $a^{\mathrm{hvp}}_{\mu}$ involves a momentum integral, knowledge of the $Q^2$ dependence of the vacuum polarization function
is needed including uncertainties and point-by-point correlations.
To estimate these from Ref.~\cite{Ce:2022eix} we computed the statistical correlation for a subset of ensembles, and assumed the systematic errors as $100\%$ correlated.  
We find a Pearson correlation coefficient of 0.8 between $a^{\mathrm{hvp}}_\mu$ and $\Delta\alpha^{(5)}(M^2_Z)$. 
The various error sources enter with $Q^2$-dependent weights that breaks the otherwise nearly perfect correlation between the two quantities.
On the other hand, the assumed correlation within each source of uncertainty is much less significant,
as even taking the systematic errors to be uncorrelated reduces the correlation merely by a few percent. 
As for the case of cross section data, Refs.~\cite{Davier:2023cyp,Malaescu:2020zuc} estimate the correlation between $a^{\mathrm{hvp}}_{\mu}$ 
and the low energy contribution ($< 2$~GeV) to $\Delta\alpha^{(5)}(M_Z)$ to $0.8$, as well.
We show these results in Fig. \ref{fig:confidenceamuda}.

\begin{figure}[t] 
    \centering
    \includegraphics[width=0.48\textwidth]{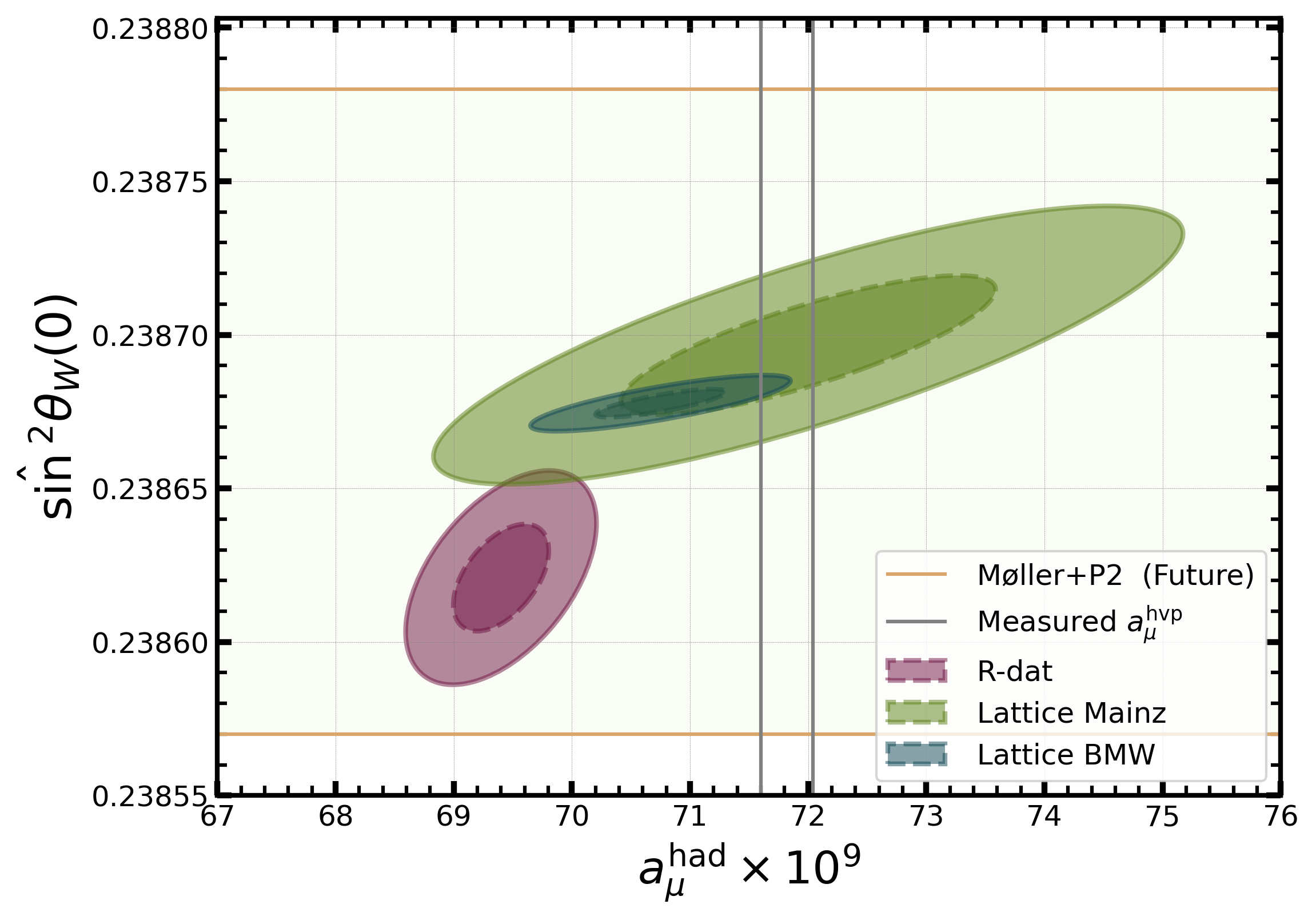}
    \caption{$\Delta\chi^2 = 1$ and $\Delta\chi^2 = 4$ contours for $\hat{s}^2(0)$ and $a^{\mathrm{hvp}}_\mu$ 
    ({\em cf.\/} Figs.~\ref{fig:confidence}) and \ref{fig:confidenceamuda}).}
    \label{fig:confidencedsinamu}
\end{figure}

Finally, the calculation of the correlation between $\hat{s}^2(0)$ and $a^{\mathrm{hvp}}_{\mu}$ from LQCD requires the point-by-point correlation of each flavor separately.
However, since the dominant errors are very similar for the $\gamma Z$ and $\gamma\gamma$ vacuum polarization functions at $\mu \approx 2$~GeV, 
one can expect the correlation of $a^{\mathrm{hvp}}_{\mu}$ with $\hat{s}^{2}(0)$ to be about the same as the one with $\Delta\alpha^{(5)}(M_Z)$. 
The result is shown in Fig.~\ref{fig:confidencedsinamu}. 

In summary, if one uses either the Mainz LQCD result~\cite{Ce:2022eix} or else cross section data as input into EW global fits, 
the values given in the upper or lower panel of Tab.~\ref{tab:MainzResults} apply, respectively.
The parameter dependencies from Eq.~(\ref{eq:parameters}) and the condensates are not included, but can easily be added by using Eq.~(\ref{eq:kappa});
for the data driven approach the corresponding dependencies can be found in Refs.~\cite{Erler:2023hyi,Erler:2017knj}.

\begin{table}[t]
\centering
\begin{tabular}{|l|r|lll|}
\hline
\rule{0pt}{2.5ex} parameter & result~~~~~~~~~~ & \multicolumn{3}{c|} {correlations} \\
\hline
\rule{0pt}{2.5ex} $\hat{\kappa}(0)-1$& ~$(323.4 \pm 1.0)\times10^{-4}$~ & ~1.0 & ~0.98 & ~0.9~ \\
\rule{0pt}{0ex} $\Delta\alpha^{(5)}(M_Z)$~ & ~$(279.4 \pm 0.9)\times10^{-4}$~ & ~0.98 & ~1.0 & ~0.8 \\
\rule{0pt}{0ex} $a^{\mathrm{hvp}}_{\mu}$  & ~$ (72.0 \pm 1.6) \times10^{-9}$~ & ~0.8 & ~0.8 & ~1.0 \\
\hline
\rule{0pt}{2.5ex} $\hat{\kappa}(0)-1$& ~$(320.1 \pm 0.8) \times 10^{-4}$~ & ~1.0 & ~0.7 & ~0.5~ \\
\rule{0pt}{0ex} $\Delta\alpha^{(5)}(M_Z)$~ & ~$(276.2 \pm 0.5) \times 10^{-4}$~ & ~0.7 & ~1.0 & ~0.8~ \\
\rule{0pt}{0ex} $a^{\mathrm{hvp}}_{\mu}$ & ~$(69.4 \pm 0.4) \times 10^{-9}$~ & ~0.5 & ~0.8 & ~1.0~ \\
\hline
\end{tabular}
\caption{Values, errors, and correlations for the running of $\sin^2\hat\theta_W$, of the hadronic contribution to $\hat\alpha(M_Z)$, and to $a_\mu$
when the input is provided by LQCD~\cite{Ce:2022eix} (upper panel) and by $e^+e^-$ data~\cite{Davier:2019can}  (lower panel).}
\label{tab:MainzResults}
\end{table}

Constraints on $\Delta\alpha^{(5)}(M_Z)$ are important for the SM prediction of the mass $M_W$ of the $W$ boson.
Inserting the values~\cite{ParticleDataGroup:2024pth} $m_t = 172.85$~GeV and $M_Z = 91.1884 $~GeV together with the Higgs boson mass,
$M_H = 125.10$~GeV and the central values of the heavy quark masses and the strong coupling constant in Eq.~(\ref{eq:parameters})
 into the numerical formula obtained in Refs.~\cite{Freitas:2002ja,Awramik:2003rn}, we can compute $M_W$ from a given value of $\Delta\alpha^{(5)}(M_Z)$. 
The results are shown in Tab.~\ref{tab:Mwpredictionsandmeasurment} together with the experimental world average~\cite{LHC-TeVMWWorkingGroup:2023zkn}, 
which excludes the recent discrepant result by the CDF Collaboration at the Tevatron~\cite{CDF:2022hxs}.

Using the correlations obtained here, we can compute the shifts in the predictions of $M_W$, when $a^{\mathrm{hvp}}_{\mu}$ is adjusted such that 
the SM prediction of $a_\mu$ in Ref.~\cite{Aoyama:2020ynm} would coincide with the experimental value~\cite{Muong-2:2023cdq}. 
In the case of the data driven approach, $a^{\mathrm{hvp}}_{\mu}$ would shift by 6~times its uncertainty, 
which for a correlation of 0.8 implies that $\Delta\alpha^{(5)}(M_Z)$ has to increase by $2\times10^{-4}$. 
This translates into a decrease in the SM prediction of $M_W$ by 4~MeV. 
On the other hand, given the good agreement of LQCD with the current experimental value, the change is only 0.2~MeV and 0.8~MeV 
for the results~\cite{Ce:2022eix} and~\cite{Budapest-Marseille-Wuppertal:2017okr}, respectively. 
A complete investigation of the impact of LQCD on EW global fits is left for future work. 

In this letter we introduced a procedure for the systematic implementation of the hadronic vacuum polarization obtained from lattice QCD to $\hat{s}^2(0)$ and other high-precision EW observables.
We used the RGE to re-sum higher order logarithms entering the calculation of $\hat{s}^2(0)$, 
and presented a parametric formula for a straightforward implementation in global EW fits. 
Furthermore, we compared this result with the one obtained from cross section data, 
and found a tension which is consistent with the known one in $\Delta\alpha^{(5)}(M_Z)$. 
Finally, we quantified the correlation of $\hat{s}^2(0)$ with $\Delta\alpha^{(5)}(M_Z)$ and $a^{\mathrm{hvp}}_{\mu}$.
As an example for the implications of the strong positive correlations that we found on other EW observables we estimated the effects 
on the SM prediction of the $W$ boson mass. 

We hope this work will help to better connect Lattice QCD with more traditional approaches to electroweak precision physics. For example, we suggest to provide the results for several $Q^2$ in a flavor-separated way, including correlations. A future direction will address the correlations introduced by the results on the strong coupling constant from LQCD.

\begin{acknowledgments}
It is a pleasure to thank Mikhail Gorshteyn, Hubert Spiesberger, and Hartmut Wittig for their invaluable comments and stimulating discussions.
We also greatly acknowledge the hospitality of the Physics Institute at UNAM in Mexico City where one of us (RFH) enjoyed a visit while completing this work. This project has received funding from the European Union's Horizon Europe research and innovation programme under the Marie Sk\l{}odowska-Curie grant agreement No 101106243.
\end{acknowledgments}

\begin{table}[t]
\centering
\begin{tabular}{|l|c|c|}
\hline
\rule{0pt}{2.5ex}  source & $M_W$ [GeV] \\
\hline
\rule{0pt}{2.5ex} $e^+e^-$ data  & ~$80.3575\pm0.0009$~ \\
\rule{0pt}{0ex} LQCD (Mainz)& ~$80.3516\pm0.0017$~ \\
\rule{0pt}{0ex} LQCD (BMW)& ~$80.3535\pm0.0005$~ \\
\hline
\rule{0pt}{2.5ex} experiment (excluding CDF)~ & ~$80.3692\pm0.0133$~ \\
\hline
\end{tabular}
\caption{Predictions for $M_W$ using the results of Ref.~\cite{Freitas:2002ja}. 
The shown errors correspond to the nonparametric ones in $\Delta\alpha^{(5)}$. 
The last line is the experimental measurement obtained from the combination in Ref.~\cite{LHC-TeVMWWorkingGroup:2023zkn}.}
\label{tab:Mwpredictionsandmeasurment}
\end{table}


\bibliography{apssamp}

\end{document}